\documentclass[%
 showpacs,
 showkeys,
 amsmath,amssymb,
]{revtex4-1}

\usepackage{graphicx}% Include figure files
\usepackage{dcolumn}% Align table columns on decimal point
\usepackage{bm}% bold math
\usepackage{CJK}

\usepackage{color}
\usepackage{amssymb}
\usepackage{amsfonts}

\usepackage[colorlinks,urlcolor=blue,linkcolor=blue,citecolor=blue,anchorcolor=blue]{hyperref}

\begin{document}

%\begin{CJK}{GBK}{song}

\preprint{APS/123-QED}

\title{Diffraction-free beams in fractional Schr\"odinger equation}% Force line breaks with \\

\author{Yiqi Zhang$^1$}
\email{zhangyiqi@mail.xjtu.edu.cn}
\author{Hua Zhong$^1$}
\author{Milivoj R. Beli\'c$^{2}$}
%\email{milivoj.belic@qatar.tamu.edu}
\author{Noor Ahmed$^1$}
\author{Yanpeng Zhang$^{1}$}
%\email{ypzhang@mail.xjtu.edu.cn}
\author{Min Xiao$^{3,4}$}
%\email{mxiao@uark.edu}
\affiliation{%
 $^1$Key Laboratory for Physical Electronics and Devices of the Ministry of Education \& Shaanxi Key Lab of Information Photonic Technique,
Xi'an Jiaotong University, Xi'an 710049, China \\
$^2$Science Program, Texas A\&M University at Qatar, P.O. Box 23874 Doha, Qatar \\
$^3$Department of Physics, University of Arkansas, Fayetteville, Arkansas 72701, USA \\
$^4$National Laboratory of Solid State Microstructures and School of Physics, Nanjing University, Nanjing 210093, China
}%

\date{\today}% It is always \today, today,

\begin{abstract}
  \noindent
  %Even in the simplest cases of beam dynamics---such as the free space propagation---the fractional Schr\"odinger equation (FSE) offers unexpected behavior.
  We investigate the propagation of one-dimensional and two-dimensional (1D, 2D) Gaussian beams in the fractional Schr\"odinger equation (FSE) without a potential, analytically and numerically.
  Without chirp, a 1D Gaussian beam splits into two nondiffracting Gaussian beams during propagation, while a 2D Gaussian beam undergoes conical diffraction.
  When a Gaussian beam carries linear chirp, the 1D beam deflects along the trajectories $z=\pm2(x-x_0)$, which are independent of the chirp.
  In the case of 2D Gaussian beam, the propagation is also deflected, but the trajectories align along the diffraction cone $z=2\sqrt{x^2+y^2}$ and the direction is determined by the chirp.
  Both 1D and 2D Gaussian beams are diffractionless and display uniform propagation.
  The nondiffracting property discovered in this model applies to other beams as well.
  Based on the nondiffracting and splitting properties, we introduce the Talbot effect of diffractionless beams in FSE.
\end{abstract}

%Use showkeys class option if keyword
\maketitle

%\tableofcontents
%
\section{Introduction}

Fractional effects, such as fractional quantum Hall effect \cite{laughlin.prl.50.1395.1983}, fractional Talbot effect \cite{wen.aop.5.83.2013},
fractional Josephson effect \cite{rokhinson.np.8.795.2012} and other effects coming from the fractional Schr\"odinger equation (FSE) \cite{zhang.prl.115.180403.2015},
introduce seminal phenomena in and open new areas of physics.
FSE, developed by Laskin \cite{laskin.pla.268.298.2000,laskin.pre.62.3135.2000,laskin.pre.66.056108.2002},
is a generalization of the Schr\"odinger equation (SE) that contains fractional Laplacian instead of the usual one,
which describes the behavior of particles with fractional spin \cite{herrmann.book.2011}.
This generalization produces nonlocal features in the wave function.
In the last decade, research on FSE was very intensive
\cite{li.jmp.46.103514.2005,dong.jmp.48.072105.2007,kowalski.pra.81.012118.2010,oliveira.jpa.44.185303.2011,lorinczi.jde.253.2846.2012,luchko.jmp.54.012111.2013,stickler.pre.88.012120.2013,zaba.jmp.55.092103.2014,tare.pa.407.43.2014}.
Even though the equivalence of SE and paraxial wave equation is known for long,
not until 2015 was the concept of FSE introduced in optics \cite{longhi.ol.36.2883.2015}, by Longhi.
In the paper, he realized a scheme to explore FSE in optics, by using aspherical optical cavities.
He has found eigenmodes of a massless harmonic oscillator,
the dual Airy functions \cite{lorinczi.jde.253.2846.2012,longhi.ol.36.2883.2015}.

Airy wave function, the eigenmode of the standard SE in free space, does not diffract during propagation.
This feature was also firstly discovered in quantum mechanics \cite{berry.ajp.47.264.1979} and then brought into optics \cite{siviloglou.ol.32.979.2007}.
In light of peculiar properties of an Airy beam, which include self-acceleration, self-healing and the absence of diffraction,
a lot of attention has been directed to accelerating diffractionless beams in the last decade.
For more information, the reader is directed to review article \cite{hu.book.2012}.
On the other hand, earlier literature on FSE mostly focused on mathematical aspects of the eigenvalue problem of different potentials,
such as the massless harmonic oscillator.
Although the harmonic potential and other potentials
\cite{naber.jmp.45.3339.2004,guo.jmp.47.082104.2006,oliveira.jmp.51.123517.2010,bay.jmp.54.012103.2013} are of high interest,
here we are not interested in the eigenvalue problem of potentials.
Rather, we focus on the dynamics of beams in FSE without any potential.
Even the simplest problem of what happens in FSE without a potential is still interesting to be explored.
Are there nondiffracting solutions? Do such solutions accelerate during propagation?
Are the solutions self-healing?
These questions are addressed in this paper.

In this article we investigate the dynamics of waves in both one-dimensional (1D) and two-dimensional (2D) FSE without a potential.
To avoid complexity of the mathematical problem of fractional derivatives, we take Gaussian beams as an example and make the analysis in the inverse space.
Methods introduced here apply to other beams as well. An approximate but accurate analytical solution to the problem is obtained,
which agrees well with the corresponding numerical simulation.
An alternative method, based on the factorization of wave equation, is also introduced.
Even though the overall analysis appears deceptively simple, the results obtained point to profound changes in the propagation of beams in FSE, as compared to the regular SE.
We discover that a Gaussian beam without chirp splits into two diffraction-free Gaussian beams in the 1D case and undergoes conical diffraction in the 2D case.
If the input Gaussian beam is chirped, it also propagates diffraction-free and exhibits uniform motion.
This uniform propagation is not much affected by the chirp.
Along the way, we also introduce the dual Talbot effect of diffractionless beams in FSE.

\section{Diffraction free beams}

The 1D FSE without potential has the form \cite{longhi.ol.36.2883.2015}
\begin{equation}\label{61eq1}
  i\frac{\partial \psi(x,z)}{\partial z}  - \frac{1}{2} \left(-\frac{\partial^2}{\partial x^2}\right)^{\alpha/2}  \psi(x,z) = 0,
\end{equation}
where $\alpha$ is the L\'{e}vy index ($1<\alpha\le2$).
When $\alpha=2$, one recovers the usual SE in free space \cite{siviloglou.ol.32.979.2007,perez-Leija.oe.21.17951.2013,eichelkraut.optica.1.268.2014}.
We consider the limiting case $\alpha=1$ \cite{zhang.prl.115.180403.2015}.
By taking Fourier transform of Eq. (\ref{61eq1}), one obtains
\begin{equation}\label{61eq2}
  i\frac{\partial \hat\psi(k,z)}{\partial z}  - \frac{1}{2} |k|  \hat\psi(k,z) = 0,
\end{equation}
with $k$ being the spatial frequency.
Equation (\ref{61eq2}) demonstrates that in the inverse space a beam propagates in a symmetric linear potential.
Recall that the potential in the inverse space is parabolic for the standard SE without potential \cite{zhang.aop.363.305.2015}.
This seemingly minor difference between the equations brings a crucial change in the behavior of beams.
The solution of Eq. (\ref{61eq1}) is easily obtained from Eq. (\ref{61eq2}) and can be written as a convolution
\begin{align}\label{61eq3}
  \psi(x,z) =& \frac{1}{2\pi} \int_{-\infty}^{+\infty} \exp\left( -\frac{i}{2}|k|z \right) \hat\psi_{\rm in}(k,0) \exp(ikx) dk \nonumber \\
  =& \frac{2}{\pi}\frac{iz}{4x^2+(iz)^2} \star \psi(x,0)
\end{align}
in which $\hat\psi_{\rm in}(k,0)$ is the Fourier transform of the input beam and $\star$ represents the convolution operation.
Notice that if $\psi(x,z)$ is a solution corresponding to $\hat\psi_{\rm in}(k,0)$,
then $\psi(-x,z)$ is a solution corresponding to $\hat\psi_{\rm in}(-k,0)$.
For spatially symmetric beams with $\hat\psi_{\rm in}(-k,0) = \hat\psi_{\rm in}(k,0)$,
the solution will also be symmetric, which can be written as $\psi(\pm|x|,z)$.
Here, we would like to note that
$
{iz}/\{\pi[4x^2+(iz)^2]\}
$
is a \textit{complex} Lorentzian function. The convolution between a Lorentzian and a Gaussian function is the Voigt function.
Because the complex Lorentzian is singular at $x=\pm z/2$ and has two peaks, the gap between the peaks will increase linearly with propagation distance.
We should note that Eq. (\ref{61eq3}) can be expressed in terms of Fox's H functions \cite{guo.jmp.47.082104.2006,bay.jmp.54.012103.2013}.
However, such a treatment does not depict the physical picture clearly, because of the complexity of mathematical expressions.
Here, we utilize an approximate analytical method to display propagation dynamics of chirped Gaussian beams.
Also, an equivalent method that provides a simple physical interpretation is presented in the Appendix.

\begin{figure*}[htbp]
	\centering
	\includegraphics[width=\textwidth]{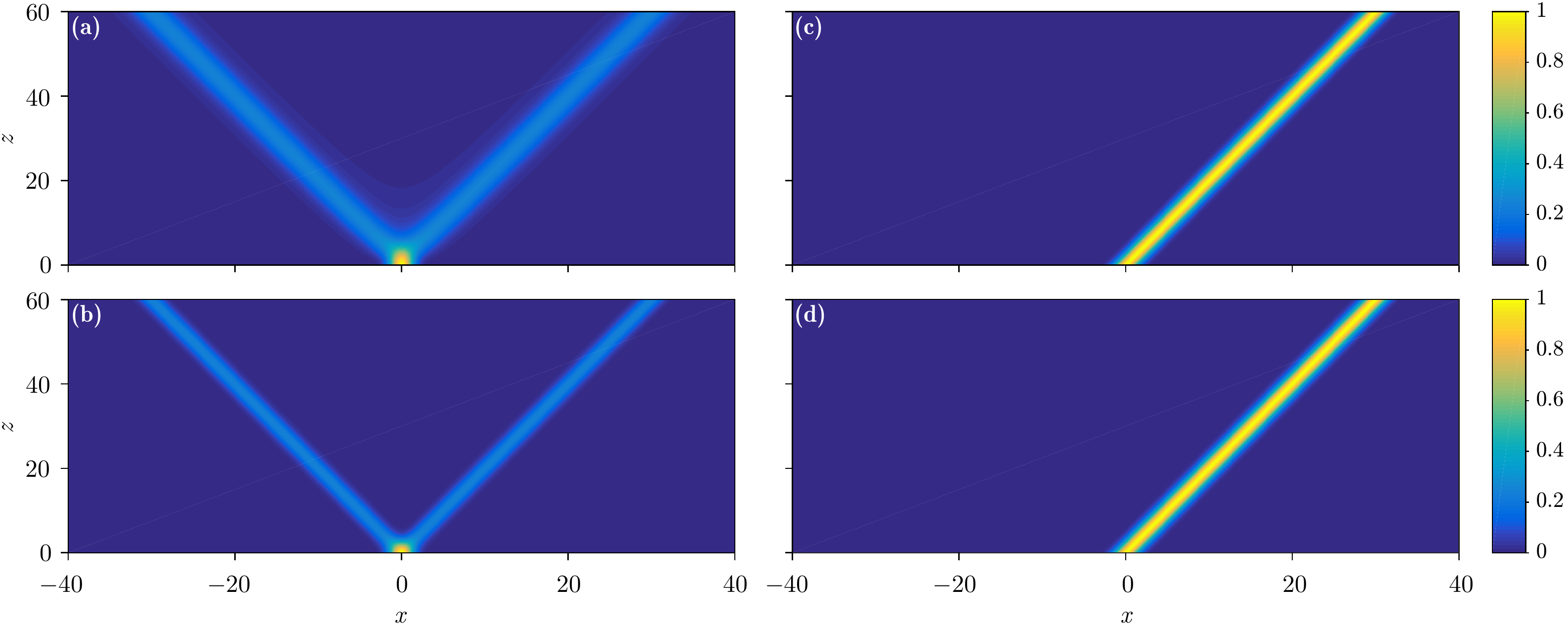}\\
	\caption{(a) Splitting of a Gaussian beam with $C=0$ during propagation in FSE.
    (b) The corresponding analytical result.
    (c) and (d) Same as (a) and (b), but for the case $C=4$.
    The Gaussian beams -- split or not -- are diffraction-free and exhibit a uniform motion.
    The parameter is $\sigma=0.25$.}
	\label{61fig1}
\end{figure*}

For simplicity, let us assume that the input is a chirped Gaussian beam
\begin{equation}\label{61eq4}
  \psi_{\rm in}(x,0)=\exp\left[-\sigma(x-x_0)^2\right]\exp(-iCx),
\end{equation}
with $x_0$ being the transverse displacement, $C$ being the linear chirp, and $\sigma$ controlling the beam width.
The corresponding Fourier transform of the input is
\begin{equation}\label{61eq5}
  {\hat\psi}_{\rm in}(k,0)=\sqrt{\frac{\pi}{\sigma}}\exp\left[-\frac{(k+C)^2}{4\sigma}\right]\exp(-ikx_0),
\end{equation}
which is also a Gaussian beam. We consider first the case $C=0$.
As is evident from the above, one cannot obtain an analytical result from the convolution directly.
Therefore, we plug Eq. (\ref{61eq5}) into Eq. (\ref{61eq3}) and after some mathematical steps,
obtain an analytical but approximate solution
\begin{align}\label{61eq6}
  \psi(x,z) \approx \frac{1}{2} \left\{ \exp\left[-\sigma \left(x_0-x+\frac{z}{2}\right)^2\right] +
  \exp\left[-\sigma\left(x_0-x-\frac{z}{2}\right)^2\right] \right\}.
\end{align}
The detailed derivation of Eq. (\ref{61eq6}) is provided in the Appendix.
From this solution one can infer that the Gaussian will split into two diffraction-free Gaussian beams -- because the beam width is not affected -- along the trajectories $z=\pm2(x-x_0)$.

The propagation of the Gaussian beam with $C=0$ is shown in Figs. \ref{61fig1}(a) and \ref{61fig1}(b),
in which Fig. \ref{61fig1}(a) depicts numerical simulation and Fig. \ref{61fig1}(b) displays the corresponding analytical result.
Clearly, they agree with each other rather well,
although the numerical solution is a bit wider than the analytical approximation.
As expected, the initial Gaussian beam splits into two diffraction-free Gaussian beams, which can be also called the one-dimensional conical diffraction.
This is starkly different from the usual behavior of Gaussian beams in free space.
The physical reason for such a splitting can be inferred from Eq. (\ref{61eq3}), which is a convolution between a Gaussian and a function with two peaks.
Since the trajectories are linear and the beam widths are unaffected during propagation, the motion of two Gaussian beams is uniform.

The solution for the case $C\neq0$ is also provided in the Appendix.
For $C$ of small absolute value,
one still observes the 1D conical diffraction, but the intensity of the two branches is not equal.
However, when $C$ is of high-enough absolute value,
the Gaussian beam depicted in Eq. (\ref{61eq5}) will be mainly in the $k>0$ or $k<0$ region,
so that the corresponding solution in Eq. (\ref{61eq1}) can be approximately written as
\begin{align}\label{61eq7}
  \psi  (x,z)=
        \exp\left[iC\left(x_0-x \pm \frac{z}{2}\right)\right]
  \exp\left[-\sigma\left(x_0-x \pm \frac{z}{2}\right)^2\right],
\end{align}
in which $\pm$ corresponds to $C\gtrless0$.
Therefore, the beam will not split during propagation, because of the high chirp.
Other than this, from Eq. (\ref{61eq7}) one can see that the chirp does not affect the trajectory or the ``velocity'' of uniform motion,
nor the transverse displacement of the beam at the output.
Thus, the chirped Gaussian beam remains diffraction-free and without acceleration during propagation.
The corresponding trajectories from numerical and analytical results are shown in Figs. \ref{61fig1}(c) and \ref{61fig1}(d).
We must emphasize that the solutions in Eqs. (\ref{61eq6}) and (\ref{61eq7}) are only approximate analytical solutions and not the exact solutions.
Still, they agree with numerical results quite well.

Since the motion of the Gaussian beam in such a model is uniform, the acceleration of caustics is missing and
the beam cannot self-heal when it encounters an obstacle (the corresponding numerical simulations not shown).
This is different from an Airy beam \cite{broky.oe.16.12880.2008, bandres.opn.24.30.2013},
a Fresnel diffraction pattern \cite{zhang.epl.104.34007.2013}, and other nonparaxial accelerating beams \cite{zhang.prl.109.193901.2012}.
It is worth mentioning that a beam always acquires a symmetric linear phase during propagation in the inverse space, according to Eq. (\ref{61eq2}),
which will not affect the beam profile except for a transverse displacement in the real space.
Thus, the nondiffracting property is feasible for all kinds of beams,
including Airy beams, Bessel-Gaussian beams, and Hermite-Laguerre-Gaussian beams, to name a few.
One should bear in mind that the quadratic phase obtained in the inverse space would change the beam profile in the real space;
this is why the Gaussian beam broadens in free space (corresponding to Eq. (\ref{61eq1}) for $\alpha=2$, which is the usual SE).

Now, it is clear that the beam splits and is diffraction-free for $\alpha=1$,
while the beam diffracts but does not split during propagation for $\alpha=2$.
Therefore, for the cases in-between, i.e., $1<\alpha<2$, the beam will both split and diffract.

\section{Talbot effect}

Considering the time-reversal symmetry of the system,
the initial Gaussian beam in Fig. \ref{61fig1}(a) can be viewed as a collision of two diffraction-free Gaussians.
From this point of view, one can consider an input composed of a superposition of equally separated Gaussian beams without chirp,
\begin{equation}\label{61eq8}
  \psi(x,0)=\sum_{n\in\mathbb{Z}} c_n \exp\left[-\sigma(x-nx_0)^2\right],
\end{equation}
with $c_n$ being the amplitude of each component.
The propagation of such a superposition is described by
\begin{align}\label{61eq9}
  \psi(x,z)= \frac{1}{2} \sum_{n\in\mathbb{Z}} c_n \left\{ \exp\left[-\sigma \left(nx_0-x+\frac{z}{2}\right)^2\right] +
  \exp\left[-\sigma\left(nx_0-x-\frac{z}{2}\right)^2\right] \right\}.
\end{align}
Let us assume that the coefficients $c_n$ are constant and independent of $n$.
If $z=2x_0$, Eq. (\ref{61eq9}) can be written as
\begin{align}\label{61eq10}
  \psi (x,z) = \frac{1}{2} c_n \left\{ \sum_{l\in\mathbb{Z}} \exp\left[-\sigma \left(lx_0-x\right)^2\right] +
  \sum_{m\in\mathbb{Z}} \exp\left[-\sigma\left(mx_0-x\right)^2\right] \right\},
\end{align}
if we let $l=n+1$ and $m=n-1$.
It is clear that Eq. (\ref{61eq10}) is reduced to Eq. (\ref{61eq8}) -- the input beam -- if we use $n$ to replace $l$ and $m$.
In other words, one can obtain a recurrent self-imaging of the input, which is known as the \textit{Talbot effect}, with the recurrence length $z_T$ known as the Talbot length.
Here, for difference, the word goes about the Talbot effect of diffractionless beams in FSE.
At $z=z_T/2$, Eq. (\ref{61eq9}) can be rewritten as
\begin{align}\label{61eq11}
  \psi(x-x_0/2,z) = \frac{1}{2} c_n  \left\{ \sum_{l\in\mathbb{Z}} \exp\left[-\sigma \left(lx_0-x\right)^2\right] +
   \sum_{n\in\mathbb{Z}} \exp\left[-\sigma\left(nx_0-x\right)^2\right] \right\},
\end{align}
which is same as the input if we substitute $l$ by $n$, except for a transverse displacement.

\begin{figure*}[htbp]
	\centering
	\includegraphics[width=\textwidth]{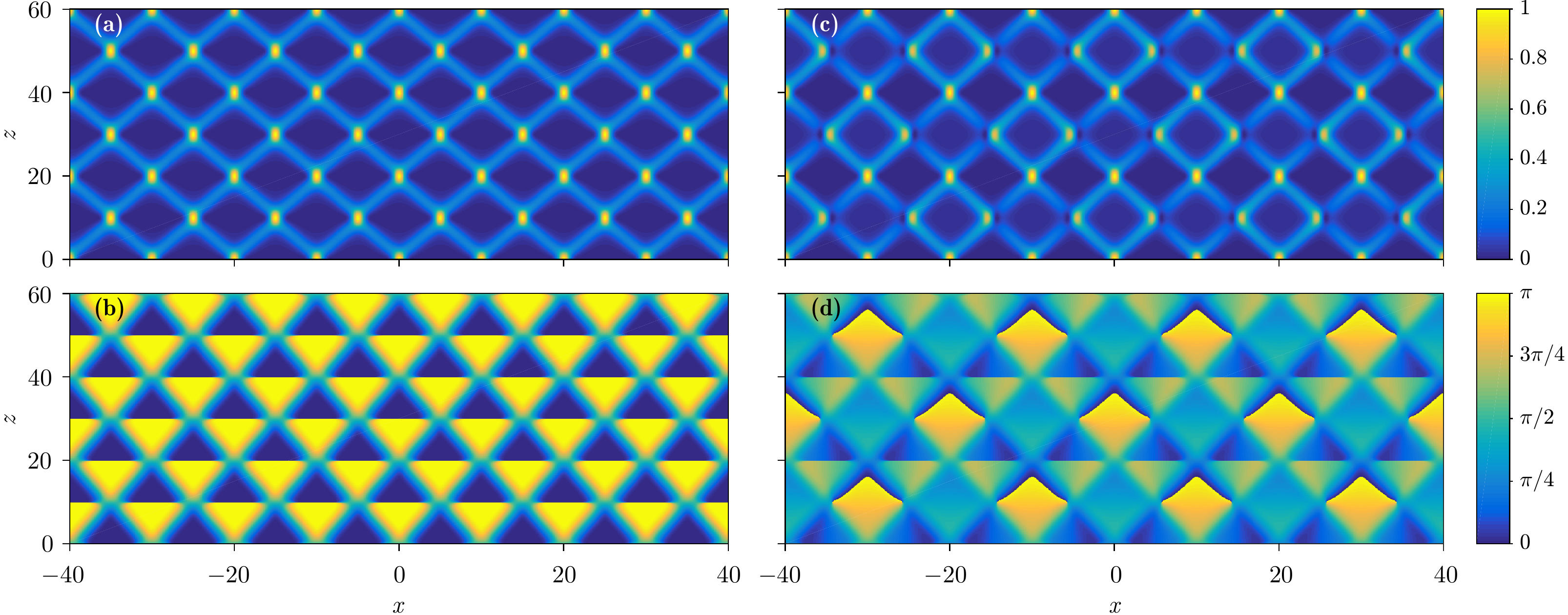}\\
	\caption{Talbot nets and carpets.
    Top panels: Intensity. Bottom panels: Phase.
    Left column: $c_n=[\cdots,1,1,1,\cdots]$. Right column: $c_n=[\cdots,i,1,i,1,i,\cdots]$.
    Other parameters: $\sigma=1$ and $x_0=10$.}
	\label{61fig2}
\end{figure*}

In Figs. \ref{61fig2}(a) and \ref{61fig2}(b), the intensity and the corresponding phase of the input beam during propagation are shown numerically,
from which one can conclude that the Talbot effect in the form of a Talbot net has formed -- the beam recovers at the Talbot length and there is a $\pi$ phase shift of the self-images at the half-Talbot length.
One should note that this Talbot effect is not the result of the diffraction of beams -- which is absent here -- but that it comes from the transverse periodicity of colliding beams \cite{lumer.prl.115.013901.2015,zhang.ol.40.5742.2015}.
The generation of such a Talbot effect can be understood as follows:
Each transverse component splits into two diffraction-free Gaussian beams,
which collide with other diffraction-free Gaussian beams that split from other components, to form the self-imaging at the half-Talbot length.
The self-imaging at half-Talbot length will then repeat itself  for another half-period, to form self-images at the Talbot length and so the full cycle is closed; it repeats, to form the full Talbot net.
Thus, the recurrence here is the consequence of peculiar input superposition and its propagation, not of the near-field diffraction.

When we choose the coefficients as $c_n=[\cdots,i,1,i,1,i,\cdots]$, the numerical results are depicted in Figs. \ref{61fig2}(c) and \ref{61fig2}(d).
One sees that the Talbot effect is still there, but the Talbot length is doubled and at the half-Talbot length the beam recovers itself, which represents the {\it dual} Talbot effect \cite{zhang.ol.40.5742.2015}.
According to the analysis in previous literature \cite{zhang.ol.40.5742.2015}, the dual Talbot effect for such a special choice of coefficients can be understood easily.

\section{Two-dimensional case}

The 2D FSE without a potential can be written as
\begin{equation}\label{61eq12}
  i\frac{\partial \psi(x,y,z)}{\partial z}  - \frac{1}{2} \left(-\frac{\partial^2}{\partial x^2}-\frac{\partial^2}{\partial y^2}\right)^{\alpha/2}  \psi(x,y,z) = 0.
\end{equation}
For $\alpha=1$, the 2D FSE in the inverse space is
\begin{equation}\label{61eq13}
  i\frac{\partial \hat\psi(k_x,k_y,z)}{\partial z}  - \frac{1}{2} \sqrt{k_x^2+k_y^2}\, \hat\psi(k_x,k_y,z) = 0.
\end{equation}
The input 2D chirped Gaussian beam can be written as
\begin{equation}\label{61eq14}
  \psi(r,0)=\exp\left(-\sigma r^2\right) \exp(iC_xx+iC_yy),
\end{equation}
where $C_x$ and $C_y$ are the chirp coefficients along the $x$ and $y$ directions, and $r=\sqrt{x^2+y^2}$.

\begin{figure}[htbp]
	\centering
	\includegraphics[width=\columnwidth]{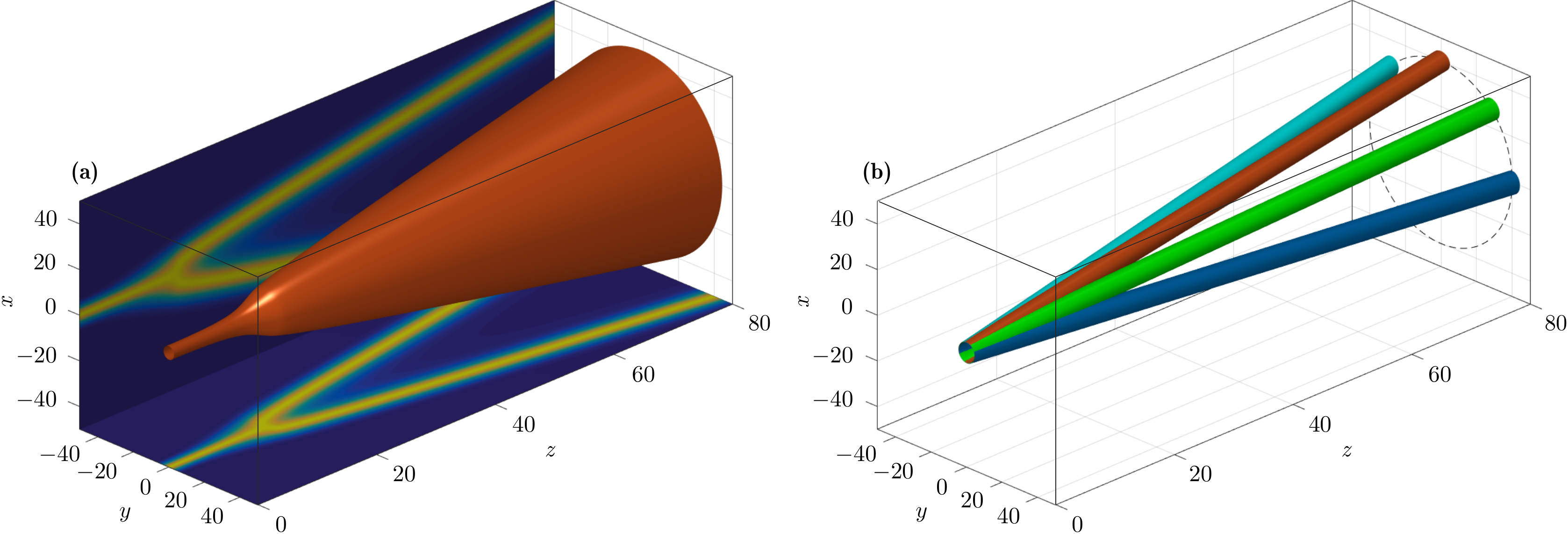}\\
	\caption{(a) Propagation of a 2D Gaussian beam without chirp, exhibiting conical diffraction.
    (b) Propagation of four chirped Gaussian beams. Clockwise from the left end, $(C_x,C_y)=(5,-5),~(5,0),~(5,5)$, and $(0,5)$, respectively.
    All Gaussians remain diffraction-free during propagation.
    The remaining parameter is $\sigma=0.04$.}
	\label{61fig3}
\end{figure}

If $C_x=C_y=0$, the Gaussian beam is not chirped, so it will undergo conical diffraction,
which is equivalent to the continuous split around the full circle.
The radius of the cone is $r=z/2$.
Such a propagation is displayed in Fig. \ref{61fig3}(a),
in which the isosurface plot depicts the panoramic view of propagation. It is indeed a conical diffraction, after an adjustment over a short propagation distance.
The intensity is normalized, to show the propagation more clearly.
The cross sections of intensity distributions along $x$ and $y$ axes are also displayed in the horizontal $x=-50$ and the vertical $y=-50$ plane.

As shown in Fig. \ref{61fig3}(b),
the chirped Gaussian beams execute uniform motion, and the direction of the ``velocity'' depends on the sign of the chirp.
One can see that no matter how the chirp coefficients change,
the output Gaussian beam will always lie on the dashed circle -- the location of the conical diffraction, as shown in Fig. \ref{61fig3}(b).
This fact confirms that the trajectory of the beam can only be along the diffraction cone.
It can also be verified that the trajectory is given by
\begin{equation}\label{61eq15}
  (x,y)=\left(\frac{z}{2}\frac{C_x}{\sqrt{C_x^2+C_y^2}},\frac{z}{2}\frac{C_y}{\sqrt{C_x^2+C_y^2}}\right),
\end{equation}
from which one finds that the chirp can only affect the location of the output beam on the dashed circle.

\begin{figure}[htbp]
	\centering
	\includegraphics[width=0.6\columnwidth]{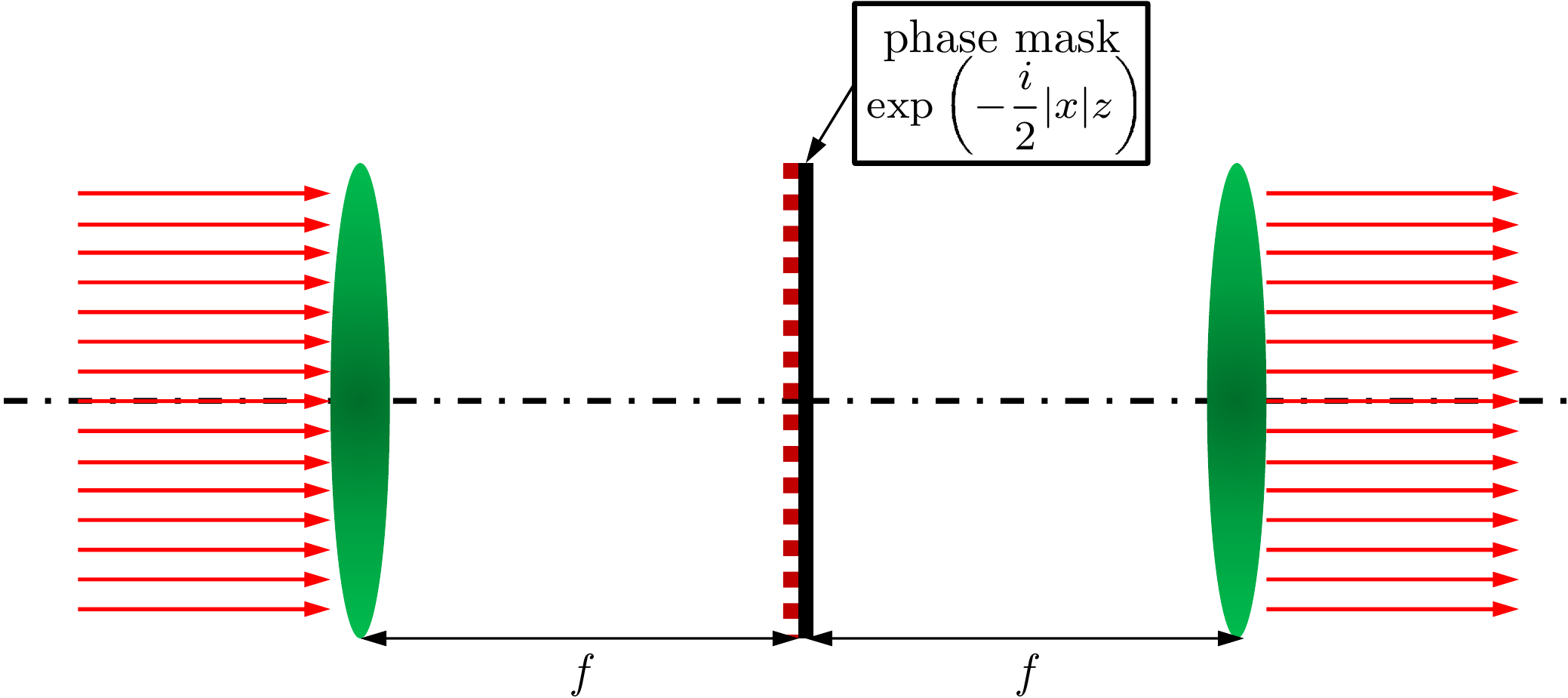}\\
	\caption{Schematics of an optical system to realize theoretical results on the free propagation in FSE. $f$ is the focal length of the convex lens.}
	\label{61fig4}
\end{figure}

For eventual experimental observation of these theoretical and numerical findings,
we propose a system composed of two convex lenses and a phase mask \cite{longhi.ol.36.2883.2015}, as shown in Fig. \ref{61fig4}.
The first lens transforms the input beam into the inverse space \cite{goodman.book.2005},
then a phase mask produces the phase change at certain propagation distance $z$, as required by Eq. (\ref{61eq2}),
and at last the second lens transforms the output beam back into the real space.
Thus, the treatment is executed in the inverse space and the management of the fractional Laplacian in Eq. (\ref{61eq1}) in the direct space is avoided.

\section{Conclusion}

In summary, we have introduced the diffraction-free beams in FSE without potential,
taking chirped Gaussian beams as an example.
The method is applicable to other beams as well.
Without chirp, a 1D Gaussian beam splits into two diffractionless Gaussian beams whose motion is uniform.
If the input is a superposition of equidistant 1D Gaussian beams, the dual Talbot effect can be realized.
We also find that conical diffraction is obtained for a 2D Gaussian beam without chirp,
while for the chirped Gaussian beams, the motion of the beams is still uniform, but the propagation direction is determined by the chirp coefficients.
Regardless of how the chirp coefficients change,
the transverse displacement is not affected, but the position of beams on the diffraction cone is.
In the end, an experimental optical implementation for such beam dynamics is proposed.
This research may not only deepen the understanding of FSE and nondiffracting beams,
but also help better control beams that show promise in various potential applications, such as producing beam splitters, beam combiners, and other.

\section*{Acknowledgements}
This work is supported by National Basic Research Program of China (2012CB921804),
National Natural Science Foundation of China (61308015, 11474228),
Key Scientific and Technological Innovation Team of Shaanxi Province (2014KCT-10),
and Qatar National Research Fund  (NPRP 6-021-1-005).
MRB also acknowledges support by the Al Sraiya Holding Group.
We thank Prof. I. M. Besieris for pointing to us the factorization of the FSE.

\renewcommand\theequation{A\arabic{equation}}
\section*{Appendix A: Derivation of Eq. (\ref{61eq6}) for the input without chirp} 

We plug Eq. (\ref{61eq5}) with $C=0$ into Eq. (\ref{61eq3}),
and try to solve for an approximate but accurate solution, as follows:
\begin{align}\label{61eq21}
  \psi(x,z) = & \frac{1}{2\pi} \int_{-\infty}^{+\infty} \exp\left( -\frac{i}{2}|k|z \right) \left[\sqrt{\frac{\pi}{\sigma}} \exp\left(-\frac{k^2}{4\sigma}\right)\exp(-ikx_0)\right] \exp(ikx) dk \nonumber \\
  = & \frac{1}{2\pi}  \sqrt{\frac{\pi}{\sigma}}
  \left\{  \int_{-\infty}^0 \exp\left(-\frac{k^2}{4\sigma}\right) \exp\left[-ik\left(x_0-x-\frac{z}{2}\right)\right] dk +
  \int_0^{+\infty} \exp\left(-\frac{k^2}{4\sigma}\right) \exp\left[-ik\left(x_0-x+\frac{z}{2}\right)\right] dk\right\}\nonumber \\
  = & \frac{1}{2\pi} \sqrt{\frac{\pi}{\sigma}}
  \left\{  \int^{+\infty}_0 \exp\left(-\frac{k^2}{4\sigma}\right) \exp\left[ik\left(x_0-x-\frac{z}{2}\right)\right] dk +
  \int_{-\infty}^0 \exp\left(-\frac{k^2}{4\sigma}\right) \exp\left[ik\left(x_0-x+\frac{z}{2}\right)\right] dk\right\} \nonumber \\
  \approx & \frac{1}{4\pi} \sqrt{\frac{\pi}{\sigma}}
  \left\{  \int^{+\infty}_{-\infty} \exp\left(-\frac{k^2}{4\sigma}\right) \exp\left[-ik\left|x_0-x-\frac{z}{2}\right|\right] dk +
  \int_{-\infty}^{+\infty} \exp\left(-\frac{k^2}{4\sigma}\right) \exp\left[-ik\left|x_0-x+\frac{z}{2}\right|\right] dk\right\}\nonumber \\
  = & \frac{1}{2} \left\{ \exp\left[-\sigma \left(x_0-x+\frac{z}{2}\right)^2\right] +
  \exp\left[-\sigma\left(x_0-x-\frac{z}{2}\right)^2\right] \right\}.
\end{align}
The approximately-equal sign in Eq. (\ref{61eq21}) is there because we presume that $x_0-x \pm z/2$ is positive for $k\gtrless0$ and negative for $k\lessgtr0$.
We should mention that in Eq. (\ref{61eq21}), the expression in the second row is obtained by reversing the sign of $k$ in the expression in the first row.
Then we sum the first two rows and obtain the expression in the third row, based on the above assumption.

\section*{Appendix B: Solution for the input with $C\neq0$} 

For this case, we rewrite $k+C=\kappa$, so according to Eq. (\ref{61eq21}) one obtains
\begin{align}\label{61eq22}
  \psi(x,z) = & \frac{1}{2\pi} \int_{-\infty}^{+\infty} \exp\left( -\frac{i}{2}|k|z \right) \left[\sqrt{\frac{\pi}{\sigma}} \exp\left(-\frac{\kappa^2}{4\sigma}\right)\exp(-ikx_0)\right] \exp(ikx) dk \nonumber \\
  = & \frac{1}{2\pi}  \sqrt{\frac{\pi}{\sigma}} \int_{-\infty}^C \exp\left(-\frac{\kappa^2}{4\sigma}\right)
      \exp\left[-i\kappa\left(x_0-x-\frac{z}{2}\right)\right] \exp\left[iC\left(x_0-x-\frac{z}{2}\right)\right] d\kappa  + \nonumber \\
    & \frac{1}{2\pi}  \sqrt{\frac{\pi}{\sigma}} \int_C^{+\infty} \exp\left(-\frac{\kappa^2}{4\sigma}\right)
      \exp\left[-i\kappa\left(x_0-x+\frac{z}{2}\right)\right] \exp\left[iC\left(x_0-x+\frac{z}{2}\right)\right] d\kappa.
\end{align}
Changing the integration limits, Eq. (\ref{61eq22}) can be rewritten as
\begin{align}\label{61eq23}
  \psi(x,z)
  = & \frac{1}{2\pi}  \sqrt{\frac{\pi}{\sigma}} \left(\int_{-\infty}^0 + \int_0^C \right) \exp\left(-\frac{\kappa^2}{4\sigma}\right)
      \exp\left[-i\kappa\left(x_0-x-\frac{z}{2}\right)\right] \exp\left[iC\left(x_0-x-\frac{z}{2}\right)\right] d\kappa  + \nonumber \\
    & \frac{1}{2\pi}  \sqrt{\frac{\pi}{\sigma}} \left(\int_0^{+\infty} - \int_0^C \right) \exp\left(-\frac{\kappa^2}{4\sigma}\right)
      \exp\left[-i\kappa\left(x_0-x+\frac{z}{2}\right)\right] \exp\left[iC\left(x_0-x+\frac{z}{2}\right)\right] d\kappa,
\end{align}
which ultimately gives the solution
\begin{align}\label{61eq24}
  \psi(x,z)
   \approx & \frac{1}{2} [1-f_+(x,z)] \exp\left[-\sigma \left(x_0-x+\frac{z}{2}\right)^2\right] \exp\left[iC\left(x_0-x+\frac{z}{2}\right)\right] + \nonumber \\
    & \frac{1}{2} [1+f_-(x,z)] \exp\left[-\sigma\left(x_0-x-\frac{z}{2}\right)^2\right] \exp\left[iC\left(x_0-x-\frac{z}{2}\right)\right],
\end{align}
where
\begin{subequations}\label{61eq25}
\begin{equation}
f_+(x,z)={\rm erf}\left[i\sqrt{\sigma}\left(x_0-x+\frac{z}{2}\right)\right] -
              {\rm erf}\left[i\sqrt{\sigma}\left(x_0-x+\frac{z}{2}\right)+ \frac{C}{2\sqrt{\sigma}} \right],
\end{equation}
and
\begin{equation}
f_-(x,z)={\rm erf}\left[i\sqrt{\sigma}\left(x_0-x-\frac{z}{2}\right)\right] -
              {\rm erf}\left[i\sqrt{\sigma}\left(x_0-x-\frac{z}{2}\right)+ \frac{C}{2\sqrt{\sigma}} \right].
\end{equation}
\end{subequations}
In Eq. (\ref{61eq25}), the error function ${\rm erf}(x)$ is defined as
\[
\operatorname{erf}(x) = \frac{2}{\sqrt\pi}\int_0^x \exp\left(-t^2\right)dt.
\]
Clearly, Eq. (\ref{61eq23}) reduces to Eq. (\ref{61eq6}) if $C=0$.
On the other hand, if $|C|$ is high enough, one of the integrals $(\int_{-\infty}^0 + \int_0^C )$ and $(\int_0^{+\infty} - \int_0^C )$
in Eq. (\ref{61eq23}) goes to 0, which leads to Eq. (\ref{61eq7}).
Similar conclusions can be also obtained based on Eqs. (\ref{61eq24}) and (\ref{61eq25}).

\section*{Appendix C: Factorization of the fractional Schr\"odinger equation in one-dimension}

We note that the analysis just described can be performed in a less rigorous manner by formally expressing the transverse fractional Laplacian in Eq. (\ref{61eq1}) for $\alpha=1$ as a {\it first-order} derivative,
\begin{equation}\label{tr}
\left(-\frac{\partial^2}{\partial x^2}\right)^{1/2}=\pm i\frac {\partial}{\partial x}.
\end{equation}
This transformation works only in the 1D case; the transverse fractional Laplacian for $\alpha=1$ in 2D cannot be
transformed into a simple combination of partial derivatives in $x$ and $y$. Still, this insight
offers an easily comprehended interpretation of the results presented in Fig. \ref{61fig1}.
Thus, when one includes Eq. (\ref{tr}) into Eq. (\ref{61eq1}), one obtains
\begin{equation}
i\frac{\partial}{\partial z}\psi(x,z)\pm\frac{i}{2} \frac{\partial}{\partial x}\psi(x,z)=0,
\end{equation}
which are the two first-order partial differential equations describing {\it advection} to the left and right. Such equations figure as {\it factor equations} in the scalar 1D wave equation,
\begin{equation}
\left(\frac{\partial}{\partial z}+ \frac{1}{2}\frac{\partial}{\partial x}\right)\left(\frac{\partial}{\partial z}- \frac{1}{2}\frac{\partial}{\partial x}\right) \psi(x,z)=0.
\end{equation}
A general solution to the wave equation
\begin{equation}
\frac{\partial^2}{\partial z^2}\psi(x,z)-\frac{1}{4}\frac{\partial^2}{\partial x^2}\psi(x,z)=0,
\end{equation}
is formally given as a sum of two arbitrary wave functions,
\begin{equation}
\psi(x,z)=\psi_1(x-z/2)+\psi_2(x+z/2),
\end{equation}
propagating in the opposite directions. This fact explains the beam propagation in Fig. \ref{61fig1}, but of course, not all such possible solutions are of interest here.

\bibliographystyle{myprx}
\bibliography{my_refs_library}

\end{document}